\documentclass[preprint,showpacs,showkeys,aps,prb]{revtex4}
\usepackage[dvips]{graphicx}

\begin{document}
\title{Flexible Macroscopic Models for Dense-Fluid
Shockwaves: \\
Partitioning Heat and Work; Delaying Stress and Heat Flux;
Two-Temperature Thermal Relaxation}
\author{
Wm. G. Hoover and C. G. Hoover                  \\
Ruby Valley Research Institute                  \\
Highway Contract 60, Box 598                    \\
Ruby Valley, Nevada 89833                       \\
}
\author{
Francisco J. Uribe                              \\
Department of Physics                           \\
Universidad Aut\'onoma Metropolitana            \\
Mexico City, Mexico 09340                       \\
}

\date{\today}

\begin{abstract}
Macroscopic models which distinguish the longitudinal and transverse
temperatures can provide improved descriptions of the microscopic shock
structures as revealed by molecular dynamics simulations.  Additionally,
we can include three relaxation times in the models, two based on Maxwell's
viscoelasticity and its Cattaneo-equation analog for heat flow,
and a third thermal, based on the Krook-Boltzmann equation.  This
approach can replicate the observed lags of stress (which lags behind
the strain rate) and heat flux (which lags behind the temperature gradient),
as well as the eventual equilibration of the two temperatures.
For profile stability the time lags cannot be too large.  By partitioning
the longitudinal and transverse contributions of work and heat and
including a tensor heat conductivity and bulk viscosity, all the
qualitative microscopic features of strong simple-fluid shockwave
structures can be reproduced.
\end{abstract}

\maketitle

\begin{figure}
\includegraphics[height=12cm,width=10cm,angle=-90]{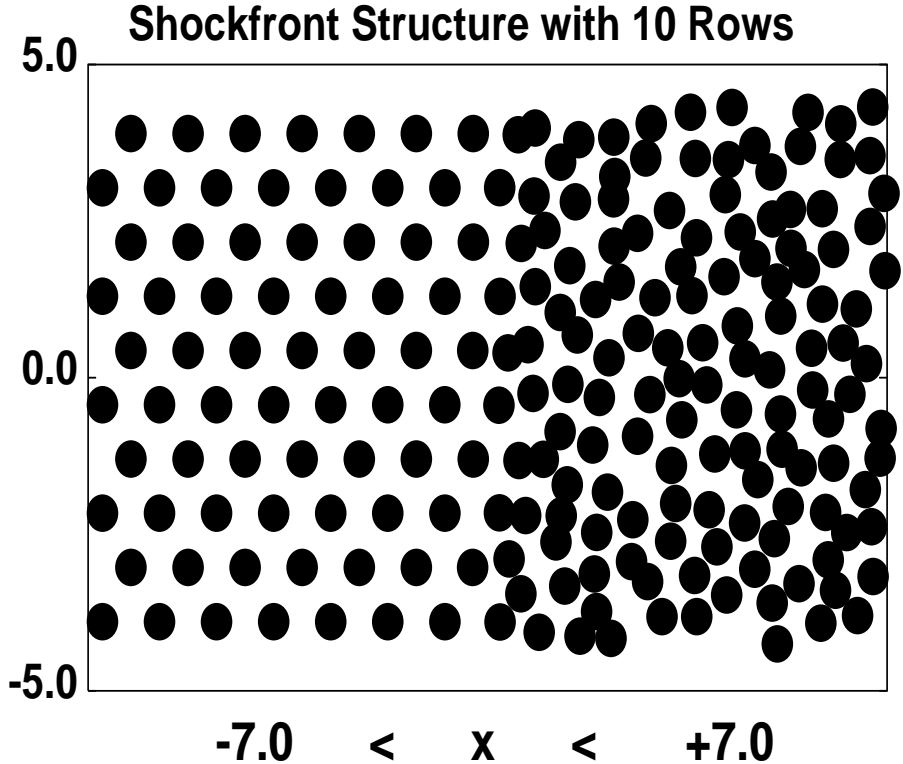}
\caption{
The steady flow shown here pictures cold material, moving to the right at
speed $u_s$ and decelerated by slower hot fluid, moving to the right at 
speed $u_s - u_p$, with $u_s$ and $u_p$ chosen to fix the shockwave
location in space, $u_{wave} = 0$.  An alternative way to generate such
shockwaves is shown in Figure 2.
}
\end{figure}

\section{Introduction: Properties of Dense-Fluid Shockwaves}

Stationary shockwaves provide the simplest possible opportunity for the study
of highly nonlinear transport in dense fluids.  In the shock-centered
steady-state coordinate frame, the nonequilibrium shock process converts
an incoming steady stream of ``cold'' material into an outgoing stream of
``hot''
fluid.  See Figure 1.  Shockwave gradients can be huge, with strainrates
in the terahertz range and correspondingly large pressure and temperature
gradients, $10^{15}$ atmospheres/centimeter and $10^{12}$
kelvins/centimeter\cite{b1}.  Despite the wildly irreversible nature of such a
nonequilibrium conversion, so long as the shock is stationary the overall
internal energy change, $E_H - E_C$, can be expressed in terms of the
equilibrium pressures and volumes of the incoming and outgoing streams of
fluid:
$$
\Delta E = E_H - E_C = (P_H + P_C)(V_C - V_H)/2 \ .
$$ 
In the steady-state coordinate frame centered on the shockwave (Figure 1),
 the incoming cold material, moving at the shock velocity $u_s$,
is decelerated to $u_s - u_p$ by the shockfront, where $u_p$ is the
``particle'', or ``piston'', velocity.

The Hugoniot relation for the energy change $\Delta E$, just given, can be
derived by eliminating the two velocities $u_s$ and $u_p$ from the three
conservation equations for mass, momentum, and energy\cite{b2}.  An
alternative shock-creation mechanism, quite practical for computer
simulation, uses the symmetric collision of two blocks of cold fluid. For
problems with a nonzero initial pressure confining pistons are required.
In either case the two blocks approach each other with velocities $\pm u_p$,
and generate two mirror-image shockwaves identical in structure to
those obtained with steady-state boundary conditions.  See again Figure 1,
as well as Figure 2, for the geometries of these two methods for generating
shockwaves.

\begin{figure}
\includegraphics[height=6in,angle=-0]{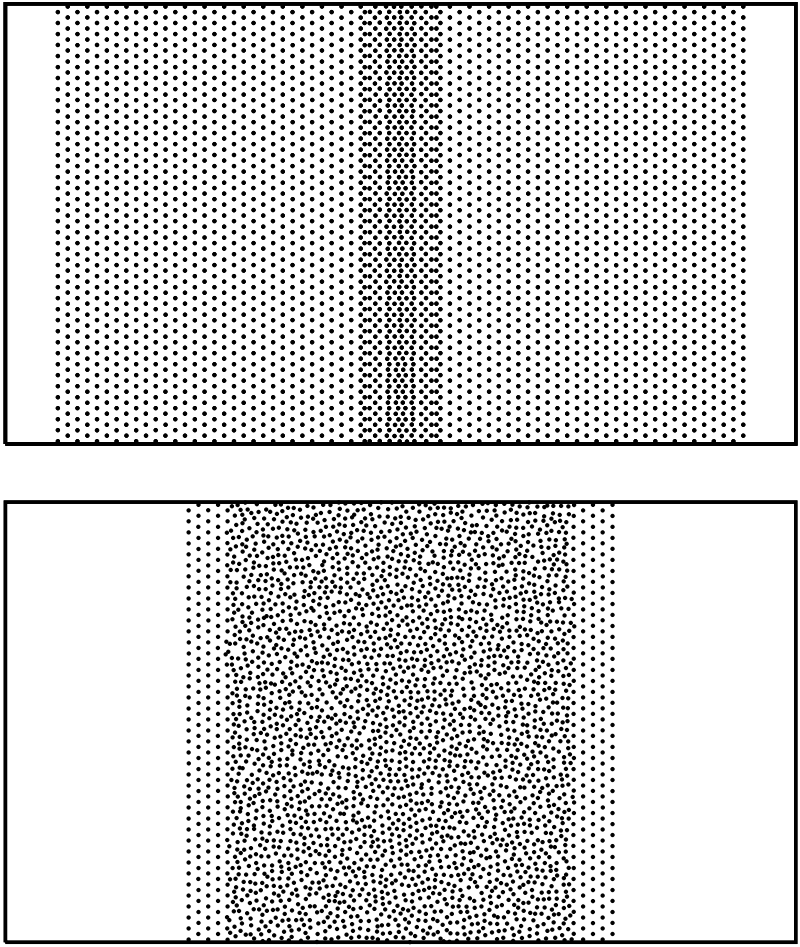}
\caption{
Here two identical blocks of zero-pressure material at $\pm u_p$ have collided
with sufficient velocity to compress the fluid to twice the initial density.
The two shockwaves at the interface between the moving cold material and the
stationary hot fluid are separating at velocities $\pm (u_s - u_p)$.  The
forces between particles, here and in Figures 1, 3, and 4, are short-ranged
repulsive forces derived from the pair potential $\phi = (10/\pi)(1-r)^3$.
}
\end{figure}

Over the last forty years a wide variety of atomistic shockwave simulations,
based on molecular dynamics, have been carried
out\cite{b3,b4,b5,b6,b7,b8,b9,b10}.  These particle-based simulations
established three interesting facts which simplify numerical treatments
of shockwaves.  First fact:  the {\em boundary
conditions} enclosing the shockwave can be implemented easily because
they are simply equilibrium states when viewed in a moving coordinate
system. See again Figures 1 and 2.  Second fact: shockwave thicknesses are
indeed only a few mean free paths\cite{b4,b5,b11}, as predicted 
for gases by numerical solutions of the Boltzmann
equation\cite{b12,b13,b14,b15,b16,b17}.  The small scale of shockwaves
makes molecular dynamics simulations relatively simple to carry out.
Third fact: one-dimensional shockwaves are stable\cite{b7}, as shown in
Figure 3.  Stability means that it is sensible to measure and compute shockwave
profiles in which density, velocity, and energy are all expressed as functions
of a single longitudinal coordinate, here chosen to define the $x$ axis.

In addition to these simplifying facts there are three more facts which
complicate rather than simplify numerical treatments.  They deserve more
discussion and form the heart of the present work: fourth fact:
temperature within the shockwave is a {\em tensor}, with
different longitudinal and transverse values.  Mott-Smith predicted
the details of this complication for gases\cite{b13}, by using an
approximate bimodal velocity distribution (a spatially-varying linear
combination of the cold and hot Maxwellian distributions).  We discuss
the meaning of ``temperature'' in the following
Section II\cite{b7,b8,b9,b12,b13,b14,b18}.

A fifth fact, discovered in the course of comparisons of atomistic
simulations with continuum predictions, is that the nonlocality of
atomistic interactions introduces an essential dependence of spatial averages
on the averaging algorithm itself.  Any continuum treatment which
aims to describe two- or three-dimensional phenomena must come to
grips with an appropriate choice of averaging algorithm.   Lucy's
one-, two-, and three-dimensional weighting functions used in smooth
particle applied mechanics\cite{b19,b20,b21} provide a particularly
appealing solution to the problem.  Averaging is addressed in Section III.

Last, a sixth fact, discovered more recently, is that relaxation and lag
are characteristic of shockwaves.  Strong shockwaves
display cause-and-effect relaxation, with the shear stress,
$\sigma = (P_{yy}-P_{xx})/2$,  responding to the strain rate
$\dot \epsilon = (du/dx)$ and the heat flux $Q_x$ responding to the temperature
gradient $\nabla T$ only after noticeable delays.  These observed delay
times are of the order of the particle-particle collision time\cite{b9}.
Lag, relaxation, and delay are addressed in Section IV.

Existing models for shockwave structure, such as the linear-transport 
Navier-Stokes equations\cite{b11,b22} or the nonlinear-transport Burnett
equations\cite{b10,b15,b16,b17,b23}, need to be improved to take these
recent shock-structure observations into account.  Delay has to be included
in the models and temperature needs to have its longitudinal and transverse
components treated separately. The present work is devoted to developing and
exploring a comprehensive description of shock dynamics and developing the
numerical techniques necessary to implement the new findings into continuum
simulations.

Following these discussions of thermal anisotropy, spatial nonlocality,
and relaxation, we introduce a well-posed continuum model incorporating
all these ideas and illustrate a numerical method for solving particular
special cases in Section V.  Section VI contains a summary of our results
and an assessment of the prospects for future progress.

\begin{figure}
\includegraphics[height=12cm,width=10cm,angle=-90]{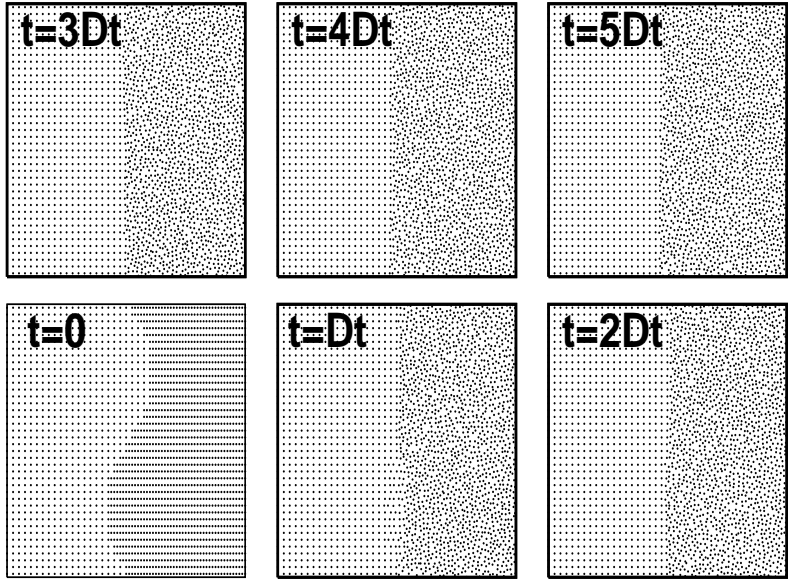}
\caption{
Six snapshots, equally spaced in time, showing the underdamped oscillation
of a sinusoidal shockwave.  The hot shocked fluid is at twice the density
of the cold unshocked material.
}
\end{figure}

\section{Kinetic Temperature and its Measurement}

Gibbs and Boltzmann related microscopic mechanics to macroscopic
thermodynamics by showing that an ideal-gas thermometer\cite{b18} satisfied the
Zeroth law of thermodynamics\cite{b24}.  Two systems at thermal equilibrium with a
Maxwell-Boltzmann ideal gas at the kinetic temperature
$$
T_{gas} = T_{eq} \equiv \langle p_x^2/mk \rangle = 
\langle p_y^2/mk \rangle = \langle p_z^2/mk \rangle \ ,
$$
are necessarily in thermal equilibrium with each other.  Thus the ideal gas
is a reliable thermometer and can be used to measure temperature in other
gases, or in liquids, or in solids.  Consider applying the ideal-gas
definition of temperature to a steady, but nonequilibrium, shockwave.  Then
there are substantial coordinate-dependent disparities between the
longitudinal and transverse kinetic temperatures,
$$
\langle p_x^2/mk \rangle = T_{xx} \ ;\ \langle p_y^2/mk \rangle = T_{yy} \ .
$$
These kinetic temperatures are velocity fluctuations about the local mean
velocity so that in the comoving measurement frame the mean values of the
momenta vanish:
$$
u(x) = \langle \dot x\rangle \ ; \
p_x \equiv   m(\dot x - \langle \dot x \rangle) =
m(\dot x - u(x)) \ ; \ \langle p_x \rangle = \langle p_y \rangle = 0 \  .
$$

The kinetic definitions for the nonequilibrium longitudinal and transverse
temperatures arise naturally if one imagines ``measuring'' them, for
particular degrees
of freedom, by putting the nonequilibrium fluid into diagnostic contact
with a comoving ideal-gas thermometer\cite{b18}.  Such a thermometer is
best thought of as a tiny sample of equilibrated gas, with the gas made
up of very many very small hard particles.  These thermometric particles
undergo impulsive collisions with selected system degrees of freedom.
If the ideal-gas particles are very small the temperature measurement
doesn't change the dynamical state of the nonequilibrium fluid\cite{b18}.
The ideal-gas nature of the thermometer makes it possible to analyze the
collisions from the two-body standpoint of the Boltzmann equation.  
Hard-disk or hard-sphere interactions between the thermometer and the
system change, on average, the total kinetic energy of a system particle
if it deviates from the thermometer's temperature.  If the thermometer
particles are instead pictured as parallel hard cubes (parallel squares
in two dimensions) with their orientations constrained, then the
temperatures $T_{xx}$ and $T_{yy}$ can be independently distinguished.

The equilibrium velocity distributions, in the thermometer, are
Maxwell-Boltzmann distributions.  Kinetic theory shows\cite{b18} that
such an ideal-gas thermometer transfers energy to/from a degree of
freedom if the kinetic energy of that degree of freedom is less/greater
than $kT_{gas}$.  When this simple mechanical definition of temperature
is used to analyze shockwave structure cause-and-effect relaxation and
thermal anisotropy are revealed.  Both these novel features need to be
tackled and described by any realistic and comprehensive shockwave model.

\section{Local Averages and their Measurement}

\begin{figure}
\includegraphics[height=12cm,width=7cm,angle=-90]{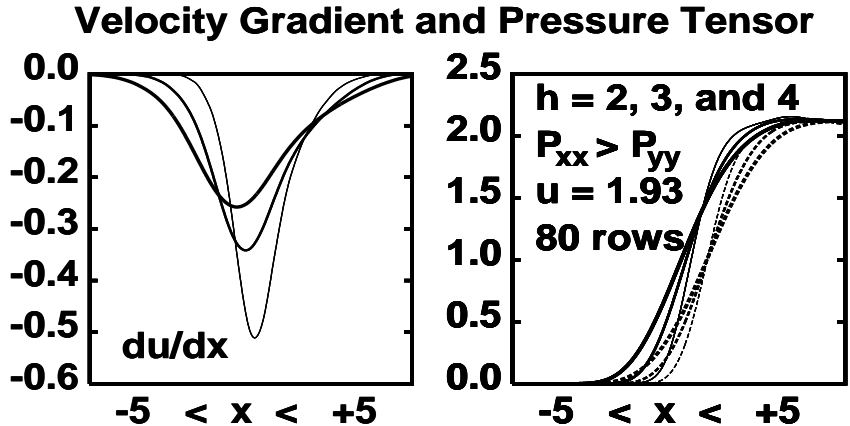}
\caption{
Dependence of the strainrate and the pressure tensor components $P_{xx}$
and $P_{yy}$ on the range of Lucy's weighting function $w(r<h)$ for the 
stationary shockwave shown in Figure 1 (but with a system width eight
times larger than that of Figure 1).  The widths of the curves in the
Figure increase with increasing $h$.
}
\end{figure}

The temperature measurements just discussed require choosing a velocity for
the thermometer.  It must be {\em comoving} with the material in order to
measure fluctuations.  But exactly what is the velocity about which the
fluctuations are measured?  A useful answer can be found based on the weight
functions used in smooth particle applied mechanics, ``SPAM''\cite{b19,b20}.
Lucy suggested that averages, at a fixed point in space, be computed using
a weight function $w(r<h)$ centered there, with an arbitrary width $h$, and
normalized, so that the integral of $w(r)$ over all space is unity.  The
simplest weight function is a polynomial chosen so that it has a smooth
maximum at the origin, $r=0$, and two continuous derivatives everywhere.  
These requirements guarantee that averages computed with the weight function,
$$
\langle F(r) \rangle \equiv \sum_j w_{rj}F_j/ \sum_j w_{rj} \ ; \
w_{rj} \equiv w(|r - r_j|) \ ,
$$
where $F(r)$ is a ``field variable'' like density, velocity, temperature,
or stress, have also two continuous spatial derivatives.  In the sums over
nearby particles $\{j\}$ it is usual to choose the range $h$ so that several
dozen particles are included.

These conditions on the weight function are sufficient to determine its
functional form:
$$
w_{Lucy}(r<h) \propto [1 - 6(r/h)^2 + 8(r/h)^3 - 3(r/h)^4] \ .
$$
Hardy's approach\cite{b21} to defining averages in shockwaves uses the same idea
as Lucy's.  Evidently $h$ must be large enough to avoid wiggles in the
resulting averages, while remaining sufficiently small for averages to
be local and inexpensive to compute.  In shockwaves a value for $h$ of
about three times the interparticle spacing is a good choice.  Figure 4
shows explicitly the dependence of the pressure tensor and the velocity
gradient averages on the range of the weight function.

When constructing continuum models designed to reproduce atomistic
simulations it is essential to specify the spatial averaging technique.
The fact that the resulting constitutive equation depends on $h$ is
simply a reminder that atomistic mechanics and continuum mechanics, though
similar, are not the same.

\section{Empirical Relaxation Models for Stress and Heat Flux}

\subsection{Maxwell's Stress Relaxation Model and its Extension to Heat Flux}

Maxwell modeled the stress relaxation characteristic of viscoelastic fluids
by introducing a stress relaxation time $\tau _\sigma $:
$$
\sigma + \tau_\sigma \dot \sigma = \eta \dot \epsilon \ .
$$
In the shockwave problem $\eta $ is the shear viscosity and $\dot \epsilon $
is the strainrate, $(du/dx)$.  Both time derivatives, indicated by the
superior dots, are comoving with the fluid.  In the absence of relaxation,
$\tau _\sigma = 0$, Maxwell's fluid model reduces to the usual Newtonian
viscous incompressible fluid, with shear stress $\sigma$ proportional  to
the instantaneous value of $\dot \epsilon$. In the absence of any imposed
strainrate ($\dot \epsilon = 0$) the initial stress decays with a
characteristic relaxation time $\tau_\sigma $.  For a delta-function strain
rate, at $t=0$, the stress has decays exponentially from its initial
value:
$$
\dot \epsilon = \delta(t=0) \longrightarrow \sigma =
(\eta /\tau)e^{-t/\tau} \ .
$$

For a relatively-simple case,  with $\eta = \tau =1$, and a localized
strain rate,
like that in the Landau-Lifshitz description of a weak shock\cite{b2}:
$$
\dot \epsilon = \frac{1}{e^{-t} + e^{+t}} \ ,
$$
Maxwell's model has an analytic solution:
$$
\sigma(t) = e^{-t}\ln\sqrt{1 + e^{+2t}} \ .
$$
Figure 5 illustrates the stress response for the Newtonian case $\tau = 0$,
and for two Maxwellian relaxation times, $\tau_\sigma = 1$ and 
$\tau_\sigma = 4$.  Recent molecular dynamics shockwave simulations have
shown that both stress and heat flux exhibit delayed responses\cite{b9}.

Exactly the same ideas can be, and have been\cite{b25}, applied to heat
flux.  If we introduce the relaxation time $\tau _Q$ into Fourier's law
for heat flow, the result is the Cattaneo equation:
$$
Q + \tau_Q \dot Q = -\kappa \nabla T \ ,
$$
and the heat flux lags behind the temperature gradient by a time of the
order of $\tau $.

On physical grounds the time derivative here is again comoving with the
fluid.  The Cattaneo equation describes heat flux and predicts its decay,
just as did Maxwell's formulation of stress decay:
$$
\nabla T \propto \frac{1}{e^{-t} + e^{+t}} \longrightarrow
Q(t) \propto -e^{-t}\ln\sqrt{1 + e^{+2t}} \ .
$$
In the following Section we illustrate how to incorporate these relaxation
effects for stress and heat flux into a simple dense-fluid shockwave model.

\subsection{Krook-Boltzmann Thermal Relaxation}

\begin{figure}
\includegraphics[height=12cm,width=7cm,angle=-90]{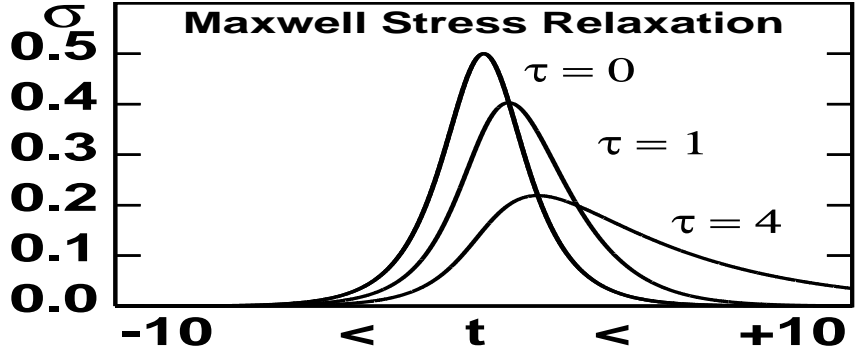}
\caption{
Delayed stress $\sigma $ in response to the strain rate $1/[e^{-t}+e^{+t}]$
with unit viscosity, $\eta = 1$.  Maxwell's relaxation time $\tau$
controls the stress response: $\sigma + \tau \dot \sigma =
\eta \dot \epsilon $.
}
\end{figure}

The Boltzmann equation\cite{b12} models the dynamics of a dilute gas in
which the gas particles undergo occasional two-body collisions.
The evolution of the velocity distribution function, $f(p,r,t)$ for the
phase-space density of particles with momentum $p$ at location $r$ at
time $t$, can be approximated by an exponential relaxation toward
equilibrium:
$$
(df/dt) \equiv [f_{eq} - f]/\tau \longleftrightarrow
f + \tau (df/dt) = f_{eq} \ .
$$
This approximate ``Krook-Boltzmann'' equation has exactly the same form
as do the Maxwell and Cattaneo relaxation equations. Here the relaxation
time $\tau$ defined by this approximate equation is of the order of the
mean collision time.  Because two-body ``conservative'' collisions
conserve mass, momentum, and energy, the equilibrium distribution,
toward which $f$ relaxes, necessarily has the same density, stream velocity,
and energy, as does the nonequilibrium distribution $f$.

Shockwaves convert macroscopic longitudinal kinetic energy into
microscopic ``thermal'' internal energy,
$$
\Delta u^2/2 \longrightarrow \Delta e  \ ,
$$
through collisions, so that it is reasonable to expect shockwave stresses
and temperatures to relax and equilibrate in a time of order the collision
time $\tau$.  We include these delay and relaxation effects in the macroscopic
model formulated in the next Section.

\section{Formulation of a Macroscopic Model}

Any solution of the continuum evolution equations,
$$
\dot \rho = - \rho \nabla \cdot u \ ; \ \rho \dot u = -\nabla \cdot P \ ; \
\rho \dot e = - \nabla u : P - \nabla \cdot Q \ ,
$$
requires constitutive models giving the pressure tensor $P$ and heat
flux $Q$ in terms of the underlying variables $\{ \ \rho, u, e \ \}$,
the density, velocity, and energy per unit mass.  For completeness,
in view of the relaxational results from molecular dynamics simulations,
we must include separate tensor temperature components, $T_{xx}$ and $T_{yy}$,
in the list of state variables.  In a gas the difference is simply related
to the pressure tensor:
$$
(T_{xx} - T_{yy}) = (P_{xx} - P_{yy})/(\rho k) \ ,
$$
where $k$ is Boltzmann's constant per unit mass.

In a dense fluid, the
potential contribution to anisotropicity is comparable to the kinetic
contribution\cite{b26}.  A semiquantitative description of the potential
part of the shear stress results if the equilibrium fluid structure is
sheared, at the strainrate $\dot \epsilon$, for the Maxwell relaxation
time $\tau _\sigma $.  The shear distortion of the pair distribution
function in dense fluids has been studied experimentally\cite{b27} and
modeled with molecular dynamics\cite{b26}.  In both cases Maxwell's relaxation
provides a good description of the potential contribution to the shear
stress.  Thus the gas-phase description of shear anisotropy must be
modified in order to describe dense fluids.

To solve this problem we choose to separate the work and heat contributing
to energy change into separate longitudinal and transverse parts.  The
simplest choice is a time-independent division of work and heat into
longitudinal and transverse parts:
$$
-\alpha\nabla u : P \longrightarrow \Delta T_{xx} \ ; \
-(1-\alpha)\nabla u : P \longrightarrow \Delta T_{yy} \ ;
$$
$$
-\beta\nabla \cdot Q \longrightarrow \Delta T_{xx} \ ; \
-(1-\beta)\nabla \cdot Q \longrightarrow \Delta T_{yy} \ .
$$
The Navier-Stokes equations correspond to the choice
$\alpha = \beta = 1/2$.  In a shockwave, where the kinetic energy is
initially longitudinal, we would expect instead
$\alpha \simeq \beta \simeq 1$.

To explore the consequences of this division we consider in what
follows a simple van der Waals model, with the energy and equilibrium
pressure expressed as sums of density-dependent and temperature-dependent
contributions.  A slightly more flexible model\cite{b8} can be based on
Gr\"uneisen's separation of the energy and pressure into corresponding
``cold'' and ``thermal'' parts.

Away from equilibrium we include Maxwell's delayed viscous
response in the pressure tensor.  For a two-dimensional fluid undergoing
uniaxial compression and with shear viscosity $\eta $ and vanishing bulk
viscosity, we have:
$$
P_{xx} = P_{\rm eq} - \sigma \ ; \ P_{yy} = P_{\rm eq} + \sigma \ ; \
\sigma + \tau_\sigma \dot \sigma =
\eta (du/dx) \ .
$$
The stress relaxation time $\tau_\sigma$ describes the delay in the
response of the shear stress $\sigma$ to the strainrate
$\dot \epsilon = (du/dx)$.

If the longitudinal and transverse temperatures are constrained to differ,
we would expect the stationary nonequilibrium heat flux vector to obey a
tensor form of Fourier's law:
$$
Q_x  = -\kappa_{xx} (dT_{xx}/dx) - \kappa_{yy} (dT_{yy}/dx) \ . 
$$
In the shockwave problem the effects of delay and eventual equilibration
both need to be included.  For simplicity we add on corresponding delays
and thermal relaxation to the continuum evolution equations for the heat
flux and the temperatures:
$$
\dot Q_x  \supset -Q_x/ \tau _Q \ ; \
\dot T_{xx} \supset (T_{yy} - T_{xx})/\tau_T \ ; \
\dot T_{yy} \supset (T_{xx} - T_{yy})/\tau_T \ .
$$

To model a dense $N$-particle van der Waals fluid, as opposed to a dilute gas,
we approximate the potential part of the thermal energy by setting
it equal to the kinetic part:
$$
E_\Phi - E_{\rm Cold} \simeq E_K = Nk(T_{xx} + T_{yy})/2 \ \longrightarrow
E_{\rm Thermal}  = Nk(T_{xx} + T_{yy}) \ .
$$

The motivation for studying such simple continuum models derives from the
results of molecular dynamics simulations of stationary
shockwaves\cite{b8,b9,b10}. Just as in the continuum case, these microscopic
molecular dynamics simulations conserve mass, momentum, and energy, so that
the stationary fluxes of these quantities,
$$
\rho u \ , \ P_{xx} + \rho u^2 \ ; \
\rho u[e +(P_{xx}/\rho) + (u^2/2)] + Q_x \ ,
$$
are constant throughout the flow.  These simulations show further that
both the stress and heat flux lag behind the strainrate and temperature
gradient.  The lags are physically reasonable from the collisional
cause-and-effect standpoint.

Newton's viscosity and Fourier's heat conduction both describe
{\em instantaneous} relationships.  Taken literally, these two
linear laws imply that the stress $\sigma $ and heat flux $Q$ respond
instantaneously, and supersonically, to the strainrate $\dot \epsilon$
and temperature gradient $\nabla T$.

Certainly such an instantaneous response is impossible.  If we imagine
reversing a time-reversible Newtonian motion of the shock process, 
another apparent shortcoming of the Navier-Stokes formulation is revealed.
Newton's and Fourier's laws,
$$
\sigma \propto \dot \epsilon \ ; \ Q_x \propto -(dT/dx) \ ,
$$
if applied to a time-reversible flow, imply that the stress changes sign
(as $\dot \epsilon$ changes sign when the motion is reversed) while the
heat flux does not (as the temperature gradient has no time-dependence).
Both conclusions are inconsistent with time-reversible Newtonian dynamics.
 
A detailed atomistic analysis of the pressure tensor and the heat flux
vector\cite{b24} shows that these functions are respectively even and odd
functions of time, so that Newton's and Fourier's ideas are necessarily
inexact as they lack the proper delay time inherent in interparticle
collisions.  Lacking a more fundamental approach to time-reversible
irreversibility, we seek to learn more by exploring explicitly the
irreversible nature of continuum models.

\section{Numerical Solutions of the van der Waals Shockwave Model}

\begin{figure}
\includegraphics[height=12cm,width=7cm,angle=-90]{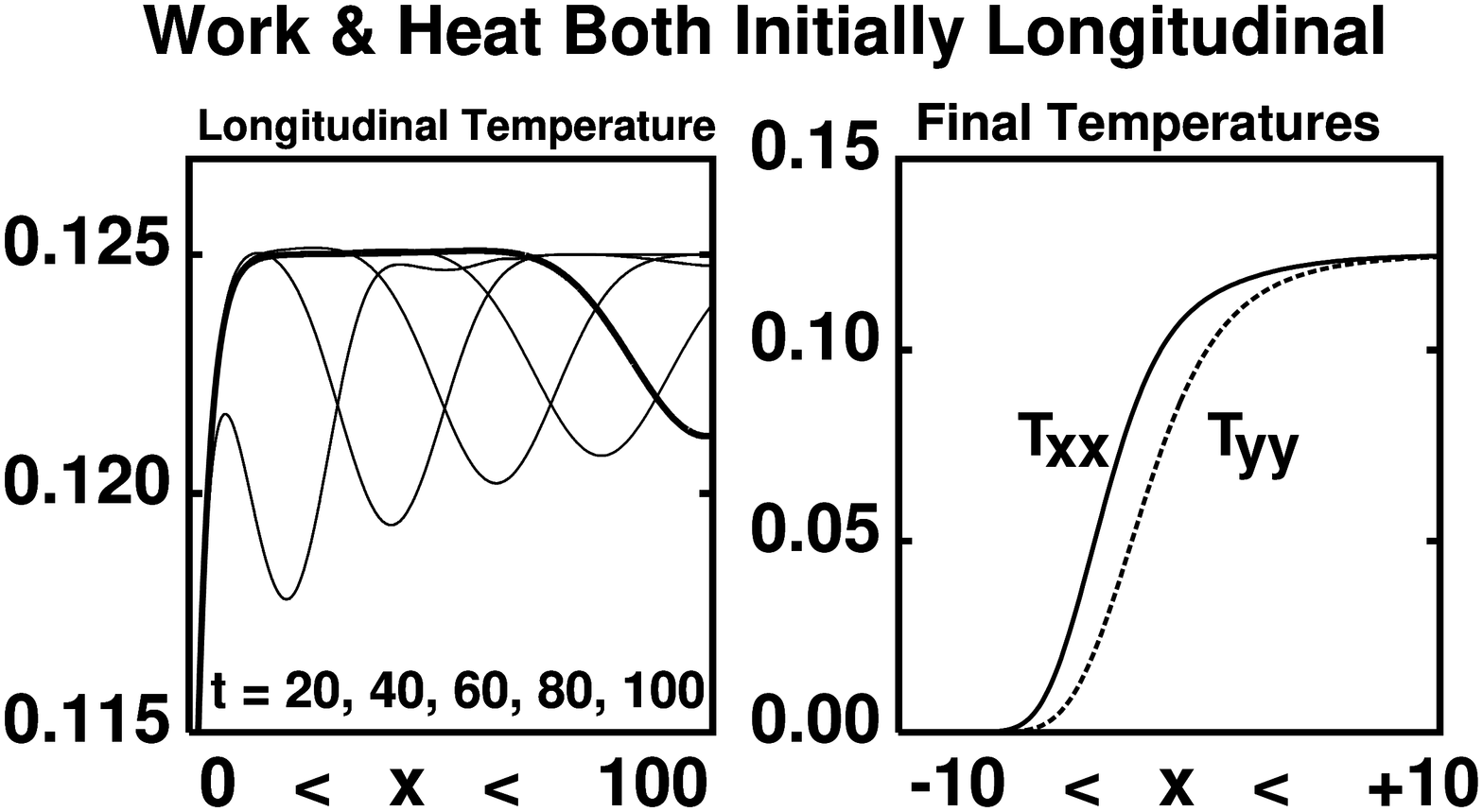}
\caption{
Development of the stationary temperature profiles in a shockwave with
shear viscosity $(\eta = 4)$, heat conductivities $\{ \kappa_{xx} =
\kappa_{yy} = 2 \}$ and relaxation times $\{ \tau _\sigma = \tau _Q =
\tau _r = 1 \} \ $ .  The converged temperatures are shown at the right.
Here the work done and the heat transfer initially affect only the
longitudinal temperature.
}
\end{figure}

A one-dimensional ``staggered grid'', with density evaluated within
$N_c$ cells of length $dx$, bounded by $N_n = N_c+1$ nodes, and with the
remaining long list of time-dependent variables
$\{ u,e,T_{xx},T_{yy},\sigma,Q\}$ given at the nodes, provides a basis
for an iterative solution of the continuum equations\cite{b10,b28}.  Our
assumption relating the energy change to the changes in the two
temperatures gives the set of nodal variables
$\{ u,T_{xx},T_{yy},\sigma,Q\}$ with the internal energy density given by
$$
e \equiv (\rho/2) + kT_{xx} + kT_{yy} \ .
$$
We have found that such an approach can be applied to
wave-structure relaxation with longitudinal and transverse work and heat
separation, as well.  The Landau-Lifshitz weak-shock solution\cite{b2} -- for
constant shear viscosity and thermal conductivity, and without any
relaxation -- makes a useful
initial condition.  In the stationary-shockwave coordinate system errors
in the initial condition move to the right, away from the shockwave, at
approximately the speed of sound.  See Figures 6 and 7 for transient results
from typical solutions of the continuum equations..

\begin{figure}
\includegraphics[height=12cm,width=7cm,angle=-90]{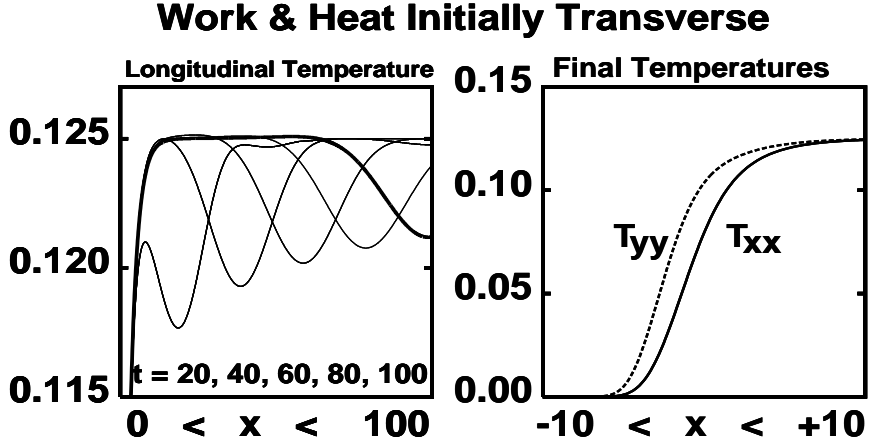}
\caption{
Development of the stationary temperature profiles in a shockwave with
shear viscosity $(\eta = 4)$, heat conductivities $\{ \kappa_{xx} =
\kappa_{yy} = 2 \}$ and relaxation times $\{ \tau _\sigma = \tau _Q =
\tau _r = 1 \} \ $.  The converged temperatures are shown at the right.
Here the work done and the heat transfer initially affect only the
transverse temperature.
}
\end{figure}

A successful numerical evolution algorithm proceeds from an initial guess
by iterating a series of four steps: (i) specify the six
dependent variables $\{ \rho_c,v_n,T_{xxn},T_{yyn},\sigma_n,Q_n \}$ at all
the interior cells and nodes; (ii) compute all the remaining variables
and all the gradients with centered sums and differences:
$$
u_c(x) = [u_n(x-dx/2) + u_n(x+dx/2)]/2 \ ; \
\rho_n(x) = [\rho_c(x-dx/2) + \rho_c(x+dx/2)]/2 \ ;
$$
$$
dT_{ii}/dx = [T_{ii}(x+dx/2)-T_{ii}(x-dx/2)]/dx \ ;
$$
(iii) compute the righthandsides of the five sets of differential
equations.

For instance, the change in energy at a particular node could be evaluated as
follows:
$$
(\partial e/\partial t)_n = -u_n(de/dx)_n -
[(P_{xx}du/dx)_n + (dQ_x/dx)_n]/\rho_n \ ;
$$
(iv) use the fourth-order Runge-Kutta method to integrate the
$N_c + 5N_n$ ordinary differential equations for one timestep $dt$,
providing the information necessary for a return to step (i) for the
execution of the next timestep.  The numerical values of the mass,
momentum, and energy, as well as their fluxes can be used to help
estimate the initial conditions.  For the twofold compression shockwave
we use to illustrate these ideas, the fluxes and boundary values are the
following:
$$
P_{eq} = \rho e \ ; \ e = (\rho /2) + T_{xx} + T_{yy} \ ;
\ e_{eq} = (\rho /2) + 2T \ ;           \
$$
$$
\rho u = 2 \ ; \ P_{xx} + \rho u^2 = 9/2 \ ; \
(\rho u)[e + (P_{xx}/\rho) + (u^2/2)] + Q_x = 6 \ ;
$$
$$
\rho: (2 \rightarrow 1) \ ; \ u: (1 \rightarrow 2) \ ; \
P: (1/2 \rightarrow 5/2) \ ; \ T: (0 \rightarrow 1/8) \ ; \
e: (1/2 \rightarrow 5/4) \ .
$$
Here the cold and hot boundary values are linked by arrows: 
$( cold \rightarrow hot)$.  Both $Q_x$ and $\sigma $ necessarily vanish
at the boundaries, $Q_x:(0 \rightarrow 0) \ ; \ \sigma :(0 \rightarrow 0)$ .

Figures 6, 7, and 8 illustrate typical solutions.  In order to circumvent
numerical instabilities in the numerical work one can (i) increase the number
of cells, (ii) reduce the timestep and/or cell size, (iii) introduce
an explicit artificial time-dependence in the parameters
$\{ \eta, \kappa, \tau \}$ in order to enhance convergence.  In this way
we have obtained solutions of the continuum shockwave model for a wide
range of parameters.  The same ideas can be used to study special cases
in which stress or heat flux are not delayed or in which temperature is
scalar rather than tensor. The sample solutions shown in Figures 6-8
show how the partition of heat and work can affect the stationary shockwave.

\begin{figure}
\includegraphics[height=12cm,width=7cm,angle=-90]{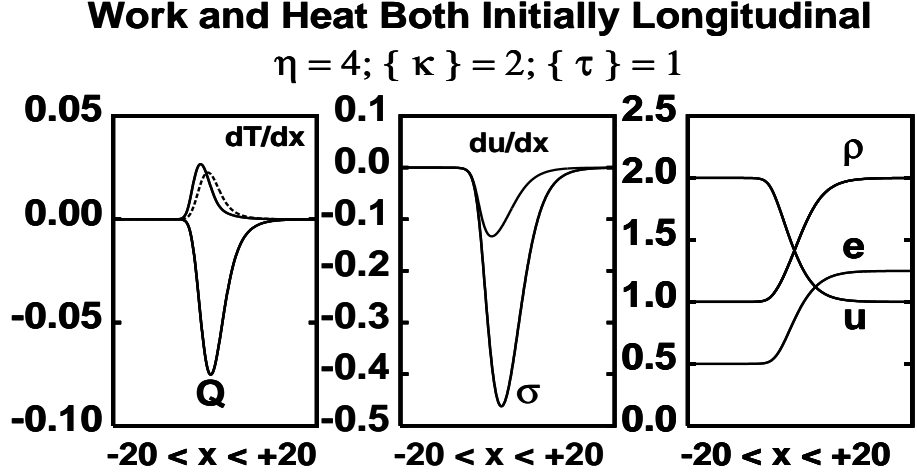}
\caption{
Stationary profiles showing the longitudinal and transverse temperature
gradients [$(dT_{xx}/dx)$ is solid; $(dT_{yy}/dx)$ is dashed] as well as
the velocity gradients.  Notice that the heat flux and shear stress lag
behind the gradients which ``cause'' them.
}
\end{figure}

\section{Conclusions and Prospects}

The simulation of nonequilibrium stationary states with molecular dynamics,
particularly in the two-block geometry of Figure 2, emphasizes Loschmidt's
reversibility paradox\cite{b12,b24,b29,b30}.  Evidently a movie of exactly
the same dynamical states, played backward in time, satisfies
all the microscopic motion equations.  Such reversed motions contradict
macroscopic physics and are never observed in practice.  They would be
 inherently Lyapunov unstable, with any small perturbation
(such as roundoff in the last place) growing exponentially in time and so
destroying the reversed trajectory.

Because time-reversed solutions of the Newtonian equations of motion are not
observable it is legitimate to use time-irreversible models in interpreting
the solutions.  The noticeable time-delays, for both stress and heat flux,
 observed in these solutions legitimates also the use of
Maxwell-Cattaneo-Krook relaxation.  These innovations are useful to the goal
of finding macroscopic descriptions conforming to microscopic observations.

Although we have been able to find stable solutions for the most general
description considered here (three relaxation times, partition of heat and work,
tensor temperature) there are stringent limits on the parameter ranges for
which such solutions exist.  On physical grounds stress and heat flux
relaxation must be relatively rapid.  A fluid's memory cannot be too long.
A systematic study of stability is complicated by the large number of
parameters involved.  Nevertheless, carefully chosen example cases should
shed additional light on the physics of relaxation and of strong shockwaves.
At the moment the step of generalizing the physical ideas further, for 
instance by considering the state dependence of the relaxation times, is
premature.  But we can confidently expect progress there in the future.

\section{Acknowledgments}  We specially thank Brad Holian for his comments,
suggestions, and prepublication copies of Reference 10.  Vitaly Kuzkin
provided the motivation for this work, through his invitation and
support for WGH and CGH's contributions to the International Summer
School--Conference ``Advanced Problems in Mechanics--2010'' organized
by the Institute for Problems in Mechanical Engineering of the Russian
Academy of Sciences in Mechanics and Engineering under the patronage of
the Russian Academy of Sciences.

\end{document}